%% file: camera-ready.tex
\documentclass{Interspeech2024}

\interspeechcameraready

\usepackage{amsmath,amsfonts}
\usepackage{algorithmic}
\usepackage{algorithm}
\usepackage{array}
\usepackage[caption=false,font=normalsize,labelfont=sf,textfont=sf]{subfig}
\usepackage{textcomp}
\usepackage{stfloats}
\usepackage{url}
\usepackage{verbatim}
\usepackage{graphicx}
\usepackage{cite}
\usepackage{cite,bm}
\usepackage{multirow}
\usepackage{amssymb}
\usepackage{comment}
\usepackage{mathtools}
\usepackage{booktabs}
\usepackage{etoolbox,siunitx}
\robustify\bfseries
\usepackage{flushend}
\usepackage{xcolor}
\usepackage{arydshln}
\usepackage{balance}

\AtBeginDocument{
  \abovedisplayskip     =5pt %
  \abovedisplayshortskip=5pt %
  \belowdisplayskip     =5pt %
  \belowdisplayshortskip=5pt %
}

\def\thineq{\hspace{-.1em}=\hspace{-.1em}}
\def\thinleq{\hspace{-.1em}\leq\hspace{-.1em}}
\def\thinneq{\hspace{-.1em}\neq\hspace{-.1em}}

\def\L{{\mathcal L}}

\def\C{{\mathbb C}}

\DeclareMathOperator*{\argmin}{arg\,min}

\title{Enhanced Reverberation as Supervision for Unsupervised Speech Separation}

\name[affiliation={1,2*}]{Kohei}{Saijo}
\name[affiliation={1}]{Gordon}{Wichern}
\name[affiliation={1}]{François}{G. Germain}
\name[affiliation={1}]{Zexu}{Pan}
\name[affiliation={1}]{Jonathan}{Le Roux}

\address{
  $^1$Mitsubishi Electric Research Laboratories (MERL), MA, USA
  $^2$Waseda University, Tokyo, Japan
  \thanks{
    *This work was done while K.\ Saijo was an intern at MERL.
   }
}
\email{saijo@pcl.cs.waseda.ac.jp, \{wichern,germain,leroux\}@merl.com}


\begin{document}

\maketitle

\begin{abstract}
Reverberation as supervision (RAS) is a framework that allows for training monaural speech separation models from multi-channel mixtures in an unsupervised manner. 
In RAS, models are trained so that sources predicted from a mixture at an input channel can be mapped to reconstruct a mixture at a target channel. 
However, stable unsupervised training has so far only been achieved in over-determined source-channel conditions, leaving the key determined case unsolved.
This work proposes enhanced RAS (ERAS) for solving this problem.
Through qualitative analysis, we found that stable training can be achieved by leveraging the loss term to alleviate the frequency-permutation problem.
Separation performance is also boosted by adding a novel loss term where separated signals mapped back to their own input mixture are used as pseudo-targets for the signals separated from other channels and mapped to the same channel.
Experimental results demonstrate high stability and performance of ERAS.

\end{abstract}

\section{Introduction}
\label{sec:introduction}
Speech separation has been intensively investigated for listening applications or as a front-end for applications such as automatic speech recognition~\cite{settle2018end, haeb2020far} or speaker diarization~\cite{kinoshita2020tackling, raj2021integration}.
Pioneered by deep clustering~\cite{dc} and permutation invariant training~\cite{dc,pit}, neural network (NN)-based approaches have become a major technique to achieve high-fidelity separation~\cite{convtasnet, tfgridnet}.
Most NN-based methods are trained in a supervised manner and rely on synthetic data as it is hard to collect pairs of mixtures and their individual sources in real environments.
NNs are however known to be vulnerable to domain mismatch, and separation models trained on synthetic data often perform poorly in real environments~\cite{zhang2021closing}.
Unsupervised speech separation techniques that can leverage recorded unlabeled mixtures can be the key to success in real-world applications~\cite{drude_unsup, togami_unsup, rccl, rccl2, munakata22_interspeech, spatial_loss, neural_fca, fast_neural_fca}.
We focus here on developing such a training technique for monaural separation.

Recently, MixIT~\cite{mixit} and remixing-based methods~\cite{remixit_journal, selfremixing, selfremixing_scratch} have shown great success for unsupervised separation.
They artificially create mixtures-of-mixtures or remix pseudo-mixtures to achieve unsupervised learning, but such artificial mixtures cause another kind of domain mismatch against the normal mixtures seen at inference.
Another direction is exploiting \textit{multi-channel} mixtures to train a monaural separation model.
Prior work has utilized spatial cues as pseudo-targets~\cite{tzinis2019unsupervised, seetharaman2019bootstrapping}, avoiding the domain mismatch issue altogether.
However, performance is bounded by the pseudo-targets' quality.

Reverberation as supervision (RAS) was proposed for effectively leveraging multi-channel mixtures to train monaural separation models~\cite{ras}.
In RAS, separated signals from an input channel are mapped to another target channel by relative room impulse response (RIR) estimation (e.g., Wiener filtering), and the model is trained to reconstruct the mixture at the target channel.
The idea is that recovering the mixture by mapping from well-separated signals is much easier than from the mixture itself, so the model will learn to separate the sources to reconstruct the mixture well.
Since both input and target are mixtures, RAS ideally overcomes both domain-mismatch and pseudo-target-quality problems.
In the original RAS~\cite{ras}, however, unsupervised learning of a two-speaker separation model using two-speaker mixtures failed, which implies that there are undesirable solutions where the model outputs signals that are not well separated but from which it is easy to recover the mixture.
Unsupervised neural speech separation leveraging over-determined mixtures (UNSSOR) has shown that we can avoid such undesirable solutions when we have more channels than sources~\cite{unssor}.
Intuitively, more constraints are imposed on the model outputs by using more microphones, because the model has to estimate signals from which the mixtures at all the microphones can be reconstructed.
Still, fully-unsupervised RAS training in the determined condition remains unsolved.

\begin{figure}[t]
\centering
\centerline{\includegraphics[width=\linewidth]{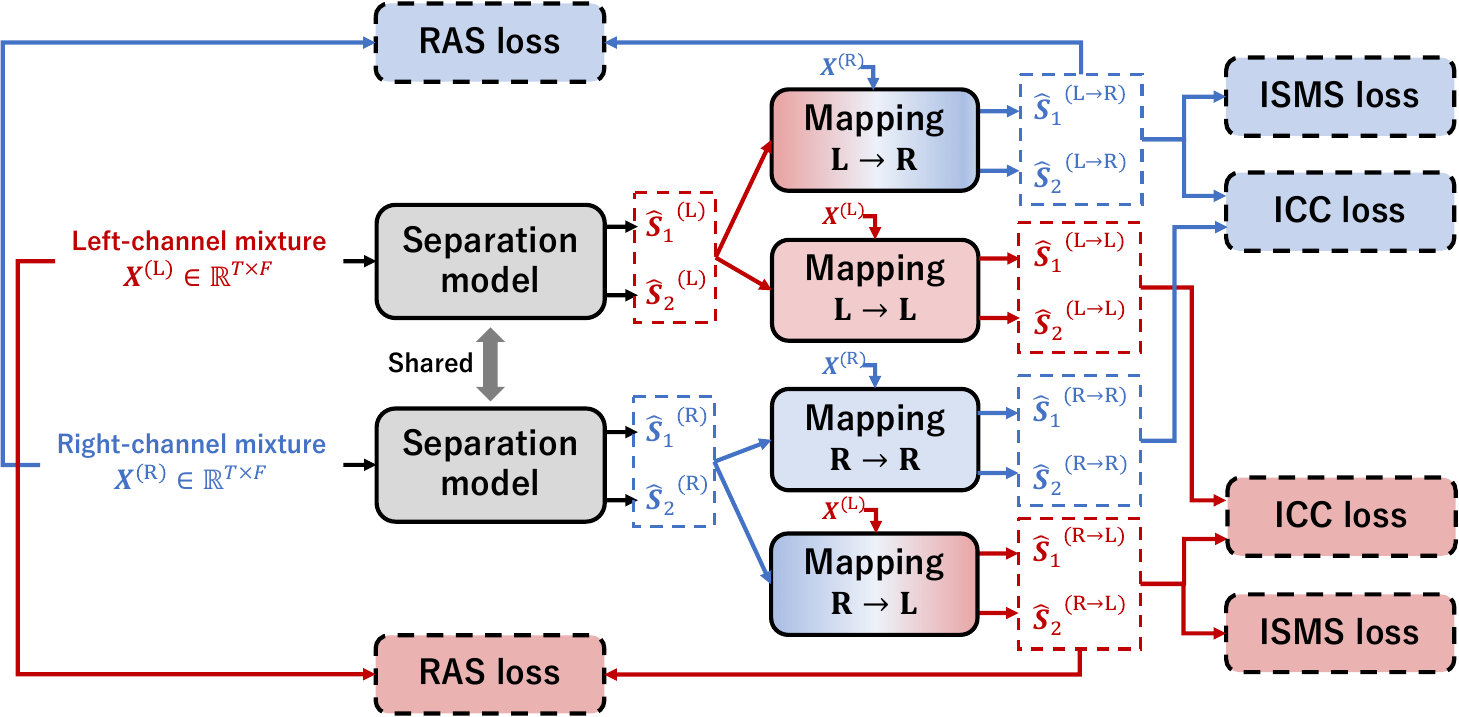}}
\vspace{-1.0mm}
\caption{
    Overview of ERAS training.
    Separated signals at the left (L) or right (R) channel are mapped to the opposite channel by relative RIR estimation, and the model is trained to reconstruct mixtures as the sum of the mapped sources (RAS loss).
    ERAS improves training stability by strongly penalize undesirable solution by ISMS loss and boosts performance by introducing an inter-channel consistency (ICC) loss aiming to make sources mapped to the same channel closer.
}
\label{fig:overview}
\vspace{-4mm}
\end{figure}

In this paper, we tackle unsupervised RAS training in the determined setup, particularly training a monaural two-speaker separation model using two-channel mixtures.
We have discovered that sources which are not well separated but are frequency-permuted result in a undesirable solution with respect to RAS loss, but such a solution can be avoided by leveraging the loss to alleviate the frequency-permutation problem~\cite{unssor}.
To boost performance, we introduce an inter-channel consistency loss in which separated signals mapped back to their input mixture are used as pseudo-targets for those separated from other channels and mapped to the same channel.
Our observation is that the former signals have higher quality and thus improve the quality of the latter.
By further introducing an effective two-stage training strategy, the proposed method, called enhanced RAS (ERAS), achieves both stable training and high separation performance.

\vspace{-.1cm}
\section{Reverberation as Supervision}
\label{sec:background}
Assuming $N$ sources are recorded with an $M$-channel microphone array, the mixture signal $x$ observed at the $m$-th microphone, $m=1,\dots,M$, can be written as
\begin{align}
  \label{eq:signal_model}
    x_{l}^{(m)} = \sum_{n=1}^{N} 
    ({h}_{n}^{(m)} \ast \bar{s}_{n})_{l},
\end{align}
where $\bar{s}_n$ denotes the dry source,
$n\thineq 1,\dots,N$ the speaker index, $l\thineq 1,\dots,L$ the time sample index, %
and $h_{n}^{(m)}$ the room impulse response (RIR) from the $n$-th source to the $m$-th microphone, 
with ${h}_{n}^{(m)} \ast \bar{s}_{n}$ being referred to as the source-image signal. We denote by $s_n^{(m)}$ the direct-path signal.
In the short-time Fourier transform (STFT) domain, we respectively denote as $X_{t,f}^{(m)}$ and $S_{n,t,f}^{(m)}$ the mixture and the $n$-th source at time frame $t\thineq 1,\dots,T$ and frequency index $f\thineq 1,\dots,F$.

\vspace{-.1cm}
\subsection{RAS and UNSSOR}
\label{ssec:ras}

RAS has been proposed to train monaural speech separation in an unsupervised manner by using multi-channel mixtures~\cite{ras}.
Specifically, RAS has been used to train a two-speaker separation model using two-channel mixtures ($M=N=2$).

Referring to the channel used as input as the reference channel and denoting its index as $m_{r}$ ($1 \leq m_r \leq M$),
the separation model $\mathcal{F}$ takes $x^{(m_r)}$ and separates it into $N$ sources:
\begin{align}
  \label{eq:separation}
    (\hat{s}_{1}^{(m_r)}, \dots, \hat{s}_{N}^{(m_r)}) = \mathcal{F}(x^{(m_r)}),
\end{align}
with $\hat{s}_{n}^{(m_r)}$ the $n$-th separated signal.
RAS aims to reconstruct the mixture at another microphone $m ~ (1 \thinleq m \thinleq M, m \thinneq m_r)$ using the separated signals.
However, the model cannot directly estimate the signals at the $m$-th channel since it does not know the position of each source and thus cannot learn to compensate for its RIR to the $m$-th channel.
RAS thus relies on an external relative RIR estimation (e.g., the Wiener filter) to map each separated signal at the $m_r$-th channel to the $m$-th channel:
\begin{align}
  \label{eq:mapping}
    \hat{s}_{n}^{(m_r\xrightarrow{}m)} = \mathcal{M}(\hat{s}_{n}^{(m_r)}, x^{(m)}),
\end{align}
where $\mathcal{M}$ denotes the mapping via relative RIR estimation, and $\hat{s}_{n}^{(m_r\xrightarrow{}m)}$ is the signal obtained by mapping $\hat{s}_{n}^{(m_r)}$ to the $m$-th channel.
Now an estimate of the mixture $x^{(m)}$ at the $m$-th microphone can be obtained by summing up the estimated sources $\hat{s}_{n}^{(m_r\xrightarrow{}m)}$ for all $n$. %
The RAS loss encourages the model to output signals from which the mixture can be reconstructed well:
\begin{align}
  \label{eq:ras_loss}
  \L_{\mathrm{RAS}}^{(m_r\xrightarrow{}m)} = \L\left(x^{(m)}, \sum\nolimits_{n}\hat{s}_{n}^{(m_r\xrightarrow{}m)}\right),
\end{align}
where $\L$ is the signal-level loss function.
Table~\ref{table:oracle_experiment} shows how well mixtures can be reconstructed from various signals at another microphone by the two filtering methods considered so far in RAS-related works, Wiener filtering~\cite{ras} and forward convolutive prediction (FCP)~\cite{fcp} (detailed in Section~\ref{ssec:mapping}).
The results show that recovering a mixture at some channel using well-separated sources is much easier than using the mixture at another channel.
Minimizing Eq.~(\ref{eq:ras_loss}) thus pushes the model to separate sources.
However, the original RAS, where $M=N=2$ is assumed, failed in a fully unsupervised setup, which implies that there are undesirable solutions where the model's outputs are not well separated but can easily reconstruct mixtures.

In~\cite{unssor}, it has been shown that such undesirable solutions in RAS can be avoided by leveraging over-determined mixtures ($M > N$).
The method, called UNSSOR, aims to reconstruct mixtures at all the $M$ channels available:
\begin{align}
  \label{eq:unssor_loss}
  \L_{\mathrm{UNSSOR}}^{(m_r)} = \sum_{m=1}^{M} \alpha^{(m)} \L_\mathrm{RAS}^{(m_r\xrightarrow{}m)},
\end{align}
where $\alpha^{(m)} (>0)$ is the loss weight of each microphone.
Unlike RAS, UNSSOR tries to recover the mixture even on the reference microphone (i.e., $m$ can be equal to $m_r$).
Since the loss $\L_\mathrm{RAS}^{(m_r\xrightarrow{}m_r)}$ has a trivial solution,  $\alpha^{(m_r)}$ is set to less than 1, while $\alpha^{(m \neq m_r)}$ is set to 1.
Intuitively, the more microphones we have, the more constraints are imposed on the model outputs because the model has to estimate the signals from which the mixtures at all microphones can be reconstructed.
Thus, having more RAS loss terms (Eq.~(\ref{eq:ras_loss})) can improve training stability, and UNSSOR can achieve unsupervised learning in over-determined conditions.
However, unsupervised RAS training in the determined condition ($M=N$) remains unsolved.

\input{tables/oracle_experiment}

\vspace{-.1cm}
\subsection{Relative RIR estimation methods for channel mapping}
\label{ssec:mapping}
In the RAS training, we have several choices of relative RIR estimation methods for the mapping $\mathcal{M}$.
The original RAS utilized the Wiener filter~\cite{ras}, while UNSSOR used FCP~\cite{fcp}.
The Wiener filtering is done in the time domain, and a source at the $m_r$-th channel is mapped to the $m$-th channel as:
\begin{align}
  \label{eq:wiener_filtering}
    \hat{s}_{n}^{(m_r\xrightarrow{}m)} = \mathcal{M}(\hat{s}_{n}^{(m_r)}, x^{(m)}) = \hat{w}_{n}^{(m)} \ast \hat{s}_{n}^{(m_r)}, \nonumber \\
    \hat{w}_{n}^{(m)} = \argmin_{w_{n}^{(m)}}  \sum_{l} |x_{l}^{(m)} - (w_{n}^{(m)} \ast \hat{s}_{n}^{(m_r)})_{l}|^2,
\end{align}
where $\hat{w}_{n}^{(m)}$ is the Wiener filter.
In contrast, FCP operates in the STFT domain.
The mapping is done with a time-invariant $K$-tap filter $\bm{g}_{n,f}^{(m)}$ (with $K_{\mathrm{past}}$ past frames, 1 current frame, and $K_{\mathrm{future}}$ future frames) as:
\begin{align}
  \label{eq:fcp}
    \hat{S}_{n,t,f}^{(m_r\xrightarrow{}m)} = (\mathcal{M}(\hat{S}_{n}^{(m_r)}, X^{(m)}))_{t,f} = (\hat{\bm{g}}_{n,f}^{(m)})^{\mathrm{H}} \tilde{\hat{\bm{S}}}_{n,t,f}^{(m_r)}, \nonumber \\
    \hat{\bm{g}}_{n,f}^{(m)} = \argmin_{\bm{g}_{n,f}^{(m)}} \sum_{t} \frac{1}{\lambda_{t,f}^{(m)}} \left|  X_{t,f}^{(m)} - (\bm{g}_{n,f}^{(m)})^{\mathrm{H}} \tilde{\hat{\bm{S}}}_{n,t,f}^{(m_r)}  \right|^2,
\end{align}
where $\tilde{\hat{\bm{S}}}_{n,t,f} \in \C^{K}$ stacks TF bins from $K_{\mathrm{past}}$ past frames and $K_{\mathrm{future}}$ future frames of $\hat{S}_{n,\cdot,f}$ with the bin from the current frame $\hat{S}_{n,t,f}$. Scaling parameter $\lambda$ is defined in~\cite{unssor} as $\lambda_{t,f}^{(m)} = (\frac{1}{M}\sum\nolimits_{m}|X_{t,f}^{(m)}|^2) + 10^{-4} \times \mathrm{max}(\frac{1}{M}\sum\nolimits_{m}|X_{t,f}^{(m)}|^2)$.

In~\cite{unssor}, it is shown that FCP leads to better performance.
This conforms with the results in Table~\ref{table:oracle_experiment}, which imply that FCP leads to higher performance and more stable RAS training.
We thus use FCP in this work\footnote{We also tried the Wiener filter~\cite{ras}: it worked but was less stable.}.
Because FCP is applied independently for each frequency, it is clearly susceptible to the frequency-permutation problem.
In~\cite{unssor}, a loss term called intra-source magnitude scattering (ISMS) loss was introduced to alleviate this problem:
\begin{align}
  \label{eq:isms_loss}
    \L_{\mathrm{ISMS}}^{(m_r\xrightarrow{}m)} = \frac{\sum\nolimits_{t} \frac{1}{N} \sum\nolimits_{n} \mathrm{var}(\mathrm{log}(|\bm{\hat{S}}_{n,t}^{(m_r\xrightarrow{}m)}|))} {\sum\nolimits_{t} \mathrm{var}(\mathrm{log}(|\bm{X}_{t}^{(m)}|))},
\end{align}
where $|\cdot|$ is magnitude and $\mathrm{var}(\cdot)$ is the variance over frequencies.
The ISMS loss favors magnitude spectrograms with smaller variance and thus mitigates the frequency-permutation problem.
The RAS loss when using FCP is replaced by:
\begin{align}
  \label{eq:ras+isms_loss}
  \L_{\mathrm{RAS+ISMS}}^{(m_r\xrightarrow{}m)} = \L_\mathrm{RAS}^{(m_r\xrightarrow{}m)} + \beta\L_\mathrm{ISMS}^{(m_r\xrightarrow{}m)},
\end{align}
where $\beta$ is a weight for the ISMS loss.

\vspace{-.1cm}
\section{Enhanced Reverberation as Supervision}
\label{sec:method}

This work tackles the unsolved problem of unsupervised learning of a monaural separation system on determined mixtures, particularly focusing on training a monaural two-speaker separation model using two-channel mixtures ($N=M=2$)\footnote{Our proposed approach algorithmically works regardless of $N$ or $M$ but we leave the experiments on other configurations as future work.}, as in the original RAS work.

\vspace{-.1cm}
\subsection{Stabilizing the training}
\label{ssec:implicit_consistency}

Interestingly, in preliminary experiments, we found that just replacing the BLSTM-based mask estimation network in the original RAS~\cite{ras} with TF-GridNet~\cite{tfgridnet} made training lead to successful separation even in the determined condition for some initial seed values (with $\beta\thineq 0.1$).
For the other seeds, it led to irrelevant solutions with very low separation quality, despite converging in terms of RAS loss.
In the failure cases, we found that the sources in the output were often frequency-permuted.

In UNSSOR, where training stability was not mentioned as an issue thanks to the over-determined setting, the ISMS loss was introduced solely to make the method frequency-permutation-solver-free, and its weight $\beta$ was set to a low value ($\beta\thineq 0.02$ for $M\thineq 6$ and $\beta\thineq 0.06$ for $M\thineq 3$)\footnote{Personal communication.}.
Since the ISMS loss was shown there to be effective for avoiding frequency-permuted outputs, we consider here leveraging it as a regularizer to stabilize training. 
Indeed, we find the ISMS loss value to be always higher when the training fails than when the training goes well. However, with weight values $\beta$ similar to those in UNSSOR or even $\beta \thineq 0.1$ as mentioned above, the training could still fail for some seed values.
We thus propose to strongly penalize such undesirable solutions by setting the ISMS loss weight to a high value such as $\beta\thineq 0.3$. 
We empirically demonstrate in Section~\ref{ssec:results} that the ISMS loss then greatly benefits the training stability.

\vspace{-.1cm}
\subsection{Inter-channel consistency loss}
\label{ssec:explicit_consistency}
While the modification in Section.~\ref{ssec:implicit_consistency} is for stabilizing the training, we propose an inter-channel consistency (ICC) loss to improve the separation performance.
Let us denote the two channels as L (left) and R (right), and define $m \in [\mathrm{L}, \mathrm{R}]$, for simplicity.
Suppose we have separated signals from both L ($\hat{s}_{n}^{(\mathrm{L})}$) and R ($\hat{s}_{n}^{(\mathrm{R})}$)\footnote{This is achieved by including both channels in the same mini-batch, i.e., the mini-batch is $[x^{(\mathrm{L})}, x^{(\mathrm{R})}, \dots]$.}, we can consider four signals from mappings L$\xrightarrow{}$R, L$\xrightarrow{}$L, R$\xrightarrow{}$R, and R$\xrightarrow{}$L (see Fig.~\ref{fig:overview}).
Then, for instance, both $\hat{s}_{n}^{(\mathrm{L\xrightarrow{}L})}$ and $\hat{s}_{n}^{(\mathrm{R\xrightarrow{}L})}$ are mapped to channel L, but we also find that $\hat{s}_{n}^{\mathrm{(L\xrightarrow{}L})}$ has better quality than $\hat{s}_{n}^{(\mathrm{R\xrightarrow{}L})}$, likely since mapping to the reference channel itself (L$\xrightarrow{}$L) is much easier than mapping to another channel (R$\xrightarrow{}$L).
This motivates us to use $\hat{s}_{n}^{\mathrm{(L\xrightarrow{}L})}$ as pseudo-label of $\hat{s}_{n}^{\mathrm{(R\xrightarrow{}L})}$ while stopping the gradient of $\hat{s}_{n}^{\mathrm{(L\xrightarrow{}L})}$, i.e., the loss term:
\begin{align}
  \label{eq:icc_loss}
  \L_{\mathrm{ICC}}^{(m_r\xrightarrow{}m)} = \frac{1}{N} \sum_{n} \L(\hat{s}_{n}^{(m\xrightarrow{}m)}, \hat{s}_{\pi}^{(m_r\xrightarrow{}m)}),
\end{align}
where $\pi$ is the optimal permutation minimizing the loss~\cite{dc, pit}.
The final combined loss is obtained with ICC loss weight $\gamma$:
\begin{align}
  \label{eq:eras+icc_loss}
  \L_{\mathrm{ERAS}}^{(m_r\xrightarrow{}m)} = \L_{\mathrm{RAS+ISMS}}^{(m_r\xrightarrow{}m)} + \gamma\L_{\mathrm{ICC}}^{(m_r\xrightarrow{}m)}.
\end{align}

\vspace{-.1cm}
\subsection{Two-stage training}
\label{ssec:isms_loss}

\input{tables/isms_loss}

While we introduced methods for stabilizing the training and improving the separation performance in Sections~\ref{ssec:implicit_consistency} and \ref{ssec:explicit_consistency}, respectively, we found that two-stage training on the ISMS loss and the ICC loss is helpful to guarantee both stability and strong performance.

\noindent \textbf{ISMS loss}:
Although the ISMS loss is used for stabilizing the training, we found that it limits the separation performance.
Table~\ref{table:isms_loss} shows the ISMS loss values computed on various signals\footnote{The input signals are randomly cut into 4-second segments to imitate the training setup. Note that when frequency-permuting the clean sources, the loss on SMS-WSJ is higher than on NF-WHAMR! because SMS-WSJ mixtures are partially overlapped. As we use 4-second segments, one of the sources is often zero-padded and frequency permutation significantly increases the variance of the log-magnitude spectrogram for zero-padded frames. Indeed, the loss on the fully-overlapped SMS-WSJ$^{\star}$ is much lower.}.
The result implies that the ISMS loss may penalize well-separated signals too much compared to other signals. 
In particular, the ISMS loss could interfere with accurate magnitude estimation because it favors spectrograms whose bins all have the same magnitude.
This analysis motivates us to improve the separation performance by removing the ISMS loss after the model gets to a certain level of separation performance.
Specifically, we first train the model with $\beta>0.0$ and then fine-tune it with $\beta=0.0$.

\noindent \textbf{ICC loss}:
In preliminary experiments, we found that the ICC loss improves the separation performance but does not contribute to stabilization of training.
We again use a two-stage approach, where we first pre-train the model with $\gamma\thineq 0.0$ for several epochs and then switch to $\gamma>0.0$.

\vspace{-.1cm}
\section{Experiments}
\label{sec:experiments}
\vspace{-.05cm}
\subsection{Experimental setup}
\label{ssec:datasets}
\vspace{-.05cm}

We use two datasets for our experiments, both contain speech from WSJ~\cite{wsj0,wsj1} and a sampling rate of 8~kHz.

\textbf{WHAMR!}~\cite{whamr} includes two-channel two-speaker mixtures (with spacing from \SIrange{15}{17}{\centi\meter}).
We use the \texttt{min} version, where the two speech signals are fully overlapped.
Reverberation times range from \SIrange{0.1}{1.0}{\second}.
Training, validation, and test sets contain 20,000 ($\sim$30h), 5,000 ($\sim$10h), and 3,000 ($\sim$5h) mixtures, respectively.
We use a noise-free version of WHAMR! (NF-WHAMR!) since the strong noise weakens our assumption that relative RIR maps signals from a channel to others, and investigation in noisy setups is left to future work.

\input{tables/results}

{\textbf{SMS-WSJ}}~\cite{smswsj} contains six-channel two-speaker mixtures.
We use the first and second microphones with a spacing of 10~cm.
The reverberation times range from \SIrange{0.2}{0.5}{\second}, and a weak Gaussian noise is added at an SNR from \SIrange{20}{30}{\decibel}.
The average overlap ratio of two speech signals is around 50\%.
Training, validation, and test sets contain 33,561 ($\sim$87.4h), 982 ($\sim$2.5h), and 1,332 ($\sim$3.4h) mixtures, respectively.

We use TF-GridNet~\cite{tfgridnet} (from ESPNet-SE~\cite{espnet, espnet_se}) as the separation model.
We set $B=4$, $D=48$, $I=4$, $J=1$ and $H=256$ (notations follow Table~1 in~\cite{tfgridnet}).
The model takes as inputs the real and imaginary (RI) parts of the mixture spectrogram and estimates the RI parts of each source.

In all the experiments, we train the model up to 100 epochs using the Adam optimizer~\cite{adam} with an initial learning rate of 1e-3.
The learning rate is halved if validation loss does not improve for two epochs.
Gradient clipping is applied with a maximum gradient $L_2$-norm of 1.
The batch size is 8 and the input mixture is 4~\si{\second} long.
Each input mixture is normalized by dividing it by its standard deviation.
For STFT, we use 
a 32~\si{\milli\second} long square-root Hann window with hop size of 8~\si{\milli\second}.
In FCP, we set $K_\mathrm{past}=19$ and $K_\mathrm{future}=1$.
As loss function $\L$, we use the $L_1$ distance between the RI components and magnitudes of the reference and the estimate, normalized by the $L_1$ norm of the input mixture, following~\cite{unssor}.

For evaluation, we use the source-image (reverberant) signals as ground truth.
We apply FCP to the network outputs to map them to the reference channel so that the ground truths and separated signals are time-aligned.
The evaluation metrics are SI-SNR, source-to-distortion ratio (SDR)~\cite{SDR-Vincent2006, fast_bss_eval}, and perceptual evaluation of speech quality (PESQ)~\cite{pesq}.

\vspace{-.1cm}
\subsection{Training stability of ERAS}
\label{ssec:results}
\vspace{-.05cm}

\input{tables/stability}

As discussed in Section~\ref{ssec:implicit_consistency}, we aim to stabilize the training by leveraging the ISMS loss.
To examine the stability of the proposed ERAS, we trained the model while changing the ISMS loss weight $\beta$, with 25 different initial seeds.
We also investigate UNSSOR-like RAS, where the RAS loss is also computed on the reference channel as in Eq.~(\ref{eq:unssor_loss}), with $\alpha^{(m_r)}\thineq 0.1$, to investigate if such a strategy is helpful for stabilizing training.
Table~\ref{table:stability} shows the number of training successes and failures when changing $\beta$, where we consider training to have failed when the average SI-SNR on the validation set does not reach 5~dB after training for 10 epochs\footnote{We believe that this definition is reasonable because the SI-SNR reaches around 10~dB after 10 epochs when the training is successful.}.
We set $\gamma\thineq 0.0$ to focus on the effectiveness of the ISMS loss.

We can clearly observe that the number of failures decreases when using higher ISMS loss weights $\beta$.
All the 25 trials succeeded on NF-WHAMR! for $\beta\thineq 0.3$ and $\alpha^{(m_r)}\thineq 0.0$.
We confirmed that $\beta=0.3$ was also an effective value on SMS-WSJ.
Although we observed that the UNSSOR-like RAS with $\alpha^{(m_r)}\thineq 0.1$ was more unstable, utilizing high $\beta$ was still effective for training stabilization.

\subsection{Performance of ERAS}
\label{ssec:results2}
In this section, we investigate performance when removing the ISMS loss and adding the proposed ICC loss in a two-stage training as described in Section~\ref{ssec:isms_loss}.
After training model \texttt{A1} with $\beta\thineq 0.3$, we use its 20-th epoch checkpoint to initialize other models, which are fine-tuned for 80 epochs afer changing loss weights as shown in Table~\ref{table:results}.
When fine-tuning, we warm up the learning rate for 4000 training steps from 0 to 1e-3.

Results of \texttt{A1} and \texttt{A2} show that removing the ISMS loss is effective, particularly on SMS-WSJ.
This is expected as the ISMS loss value on SMS-WSJ can be high even with perfect separation (cf.\ Section~\ref{ssec:isms_loss}).
On NF-WHAMR!, PESQ, which favors more accurate magnitude estimation, showed large improvements, in line with the fact that the ISMS loss leads to inaccurate magnitude estimation as discussed in Section~\ref{ssec:isms_loss}.

The result of \texttt{A3} shows that the ICC loss is also beneficial. %
Removing the ISMS loss while adding the ICC loss (\texttt{A4}) leads to the best performance.
While there is still a gap with supervised learning, the proposed ERAS improved both training stability and separation quality.

\section{Conclusion}
\label{sec:conclusion}
We proposed ERAS, a stable and high-performing unsupervised speech separation method for training a monaural separation model using multi-channel mixtures.
Inspired by the analysis that implies the ISMS loss could be effective as the regularizer to avoid undesirable solutions, we achieved stable training by setting the ISMS-loss weight high.
We also proposed ICC loss and introduced a two-stage training strategy to ensure both high stability and performance.
In future work, we will further improve ERAS by imposing more constraints such as spatial consistency~\cite{spatial_loss, falcon2023location}. Our source code is available online%
\footnote{
\url{https://github.com/merlresearch/reverberation-as-supervision}
}.

\clearpage
\balance
\bibliographystyle{IEEEtran}
\bibliography{main}

\end{document}

%% file: tables/oracle_experiment.tex
\begin{table}[t]
\centering
\sisetup{
detect-weight, %
mode=text, %
tight-spacing=true,
round-mode=places,
round-precision=1,
table-format=2.1,
table-number-alignment=center
}
\caption{
    SI-SNR~[dB] when predicting a mixture at a channel from signals at another channel with Wiener filter or FCP.
}
\vspace{-3mm}
\label{table:oracle_experiment}
\resizebox{1.0\linewidth}{!}{
\begin{tabular}{l*{4}{S}}
\toprule
& \multicolumn{2}{c}{NF-WHAMR!} & \multicolumn{2}{c}{SMS-WSJ} \\

\cmidrule(lr){2-3}\cmidrule(lr){4-5}

{Filtering input} & {Wiener} & {FCP} & {Wiener} & {FCP}  \\

\midrule
    Mixture $\downarrow$ &6.2 &\bfseries 4.3 &6.8 &\bfseries 5.1  \\ 
    
    Source-image signals $\uparrow$ &12.1 &\bfseries 13.4 &11.2 &\bfseries 11.4  \\

    Direct-path + early reflections $\uparrow$ &12.7 &\bfseries 14.4 &12.2 &\bfseries 13.1 \\ 

    Dry sources $\uparrow$ &14.2 &\bfseries 17.0 &15.2 &\bfseries 17.4  \\ 
     
\bottomrule

\end{tabular}
}
\vspace{-4mm}
\end{table}

%% file: tables/isms_loss.tex
\begin{table}[t]
\centering
\sisetup{
detect-weight, %
mode=text, %
tight-spacing=true,
round-mode=places,
round-precision=2,
table-format=1.2,
table-number-alignment=center
}
\caption{
    Oracle ISMS loss value (Eq.~(\ref{eq:isms_loss})) on validation set. In SMS-WSJ$^\star$, we only consider fully-overlapped segments.
}
\vspace{-3mm}
\label{table:isms_loss}
\resizebox{\linewidth}{!}{
\begin{tabular}{lSS[table-format=2.2]S}
\toprule
& {NF-WHAMR!} & {SMS-WSJ} & {SMS-WSJ$^\star$} \\

\midrule

    Mixture, mixture &1.00 &1.00 &1.00  \\ 
    
    Mixture, zero signal &0.50 &0.50 &0.50  \\ 

    Zero signal, zero signal &0.00 &0.00 &0.00  \\ 

    Source-image signals &1.05 &1.53 &1.61  \\ 

    Freq. permuted clean signals  &1.45 &10.83 &2.22  \\ 
     
\bottomrule

\end{tabular}
}
\vspace{-4mm}
\end{table}

%% file: tables/results.tex
\begin{table*}[t]
\centering
\sisetup{
detect-weight, %
mode=text, %
tight-spacing=true,
round-mode=places,
round-precision=1,
table-format=2.1,
table-number-alignment=center
}
\caption{
    Effectiveness of adding ICC loss and removing ISMS loss in ERAS training. $\beta$ and $\gamma$ are the ISMS loss and ICC loss weights.
}
\vspace{-3mm}
\label{table:results}
\resizebox{0.70\linewidth}{!}{
\begin{tabular}{cl*{2}{S[table-format=1.1]}*{2}{S}S[table-format=1.2,round-precision=2]*{2}{S}S[table-format=1.2,round-precision=2]}

\toprule

& & & & \multicolumn{3}{c}{NF-WHAMR!} & \multicolumn{3}{c}{SMS-WSJ} \\
\cmidrule(lr){5-7}\cmidrule(lr){8-10}
ID &Method &$\beta$ &$\gamma$ &{SI-SNR} &{SDR} &{PESQ} &{SI-SNR} &{SDR} &{PESQ} \\

\midrule

\texttt{A1} &{ERAS} &0.3 &0.0 &13.1 &13.9 &3.27 &11.3 &12.0 &2.96 \\
\texttt{A2} &{\texttt{A1@20epochs} $\xrightarrow{}$ ERAS} &\bfseries 0.0 &0.0 &13.8 &14.7 &3.65 &13.7 &14.4 &3.54 \\
\texttt{A3} &{\texttt{A1@20epochs} $\xrightarrow{}$ ERAS} &0.3 &\bfseries 0.1 &13.2 &13.9 &3.32 &11.7 &12.4 &3.07 \\
\texttt{A4} &{\texttt{A1@20epochs} $\xrightarrow{}$ ERAS} &\bfseries 0.0 &\bfseries 0.1 &\bfseries 14.7 &\bfseries 15.6 &\bfseries 3.75 &\bfseries 13.9 &\bfseries 14.5 &\bfseries 3.55 \\
\midrule
 &{Supervised} &{-} &{-} &15.7 &16.6 &3.87 &15.8 &16.4 &3.89 \\

\bottomrule

\end{tabular}
}
\vspace{-4mm}
\end{table*}

%% file: tables/stability.tex
\begin{table}[t]
\centering
\sisetup{
detect-weight, %
mode=text, %
tight-spacing=true,
round-mode=places,
round-precision=1,
table-format=2.1,
table-number-alignment=center
}
\caption{
    Number of training successes / failures among 25 trials with different initial seed values, when changing ISMS loss weight $\beta$.
    $\alpha^{(m_r)}$ is loss weight on reference microphone.
}
\vspace{-3mm}
\label{table:stability}
\resizebox{1.0\linewidth}{!}{

\begin{tabular}{lccccc}
\toprule

&\multicolumn{4}{c}{NF-WHAMR!} & {SMS-WSJ} \\
\cmidrule(lr){2-5}\cmidrule(lr){6-6}
$\alpha^{(m_r)}$ &{$\beta=0.05$} &{$\beta=0.1$} &{$\beta=0.2$} &{$\beta=0.3$} &{$\beta=0.3$} \\

\midrule

0.0 &15 / 10 &20 / \phantom{0}5 & 21 / \phantom{0}4 &\bfseries{25 / 0} &\bfseries{25 / 0} \\
0.1 &\phantom{0}0 / 25 &\phantom{0}0 / 25 &\phantom{0}9 / 16 &19 / 6 &21 / 4 \\

\bottomrule
\end{tabular}
}
\vspace{-4mm}
\end{table}